\documentclass[ras]{agutex}

\usepackage[dvips]{graphicx}

\authorrunninghead{HELMBOLDT ET. AL}

\titlerunninghead{HIGH-PRECISION TEC GRADIENTS WITH THE VLA}

\authoraddr{J.\ F.\ Helmboldt, US Naval Research Laboratory, 
Code 7213, 4555 Overlook Ave. SW, Washington, DC 20375 
(joe.helmboldt@nrl.navy.mil)}

\authoraddr{T.\ J.\ W.\ Lazio, Jet Propulsion Laboratory, California Institute of Technology, M/S 138-308, 4800 Oak Grove Dr., Pasadena, CA 91106
(joseph.lazio@jpl.nasa.gov)}

\authoraddr{H.\ T.\ Intema, National Radio Astronomy Observatory, 
520 Edgemont Rd, Charlottesville, VA 22903 
(hintema@nrao.edu)}

\authoraddr{K.\ F.\ Dymond, US Naval Research Laboratory, 
Code 7655, 4555 Overlook Ave. SW, Washington, DC 20375 
(kenneth.dymond@nrl.navy.mil)}

\begin{document}

\title{High-precision Measurements of Ionospheric TEC Gradients with the Very Large Array VHF System}

\authors{J. F. Helmboldt, \altaffilmark{1} T. J. W. Lazio, \altaffilmark{2}  H. T. Intema, \altaffilmark{3} \& K. F. Dymond, \altaffilmark{1}}

\altaffiltext{1}{US Naval Research Laboratory, Washington, DC, USA.}
\altaffiltext{2}{Jet Propulsion Laboratory, California Institute of Technology, Pasadena, CA, USA.}
\altaffiltext{3}{Jansky Fellow of the National Radio Astronomy Observatory, Charlottesville, VA, USA.}

\begin{abstract}
We have used a relatively long, contiguous VHF observation of a bright cosmic 
radio source (Cygnus A) with the Very Large Array (VLA) to demonstrate the 
capability of this instrument to study the ionosphere.  This interferometer, and others like it, can observe ionospheric 
total electron content (TEC)  fluctuations on a much wider range of scales than is possible with many 
other instruments.  
We have shown that with a bright source, the VLA can measure differential TEC values between pairs of antennas ($\delta \mbox{TEC}$) with an precision of $3 \times 10^{-4}$ TECU.  Here, we detail the data reduction and processing techniques used to achieve this level of precision.  In addition, we demonstrate techniques for exploiting these high-precision $\delta \mbox{TEC}$ measurements to compute the TEC gradient observed by the array as well as small-scale fluctuations within the TEC gradient surface.  A companion paper details specialized spectral analysis techniques used to characterize the properties of wave-like fluctuations within this data. 
\end{abstract}

\begin{article}

\section{Introduction}
The effects of the ionosphere have always been an obstacle for ground-based radio 
frequency observations of astronomical sources.  This is especially true for 
interferometric observations used to make images of objects with relatively 
small angular sizes.  VHF and UHF interferometers such as the Very Large Array 
(VLA) in New Mexico, 
the Westerbork Synthesis Radio Telescope (WSRT) in the Netherlands, the Giant 
Metrewave Radio Telescope (GMRT) in India, and the Australia Telescope Compact 
Array (ATCA), among others, are all affected by the ionosphere in the same 
way.\par
The fundamental principles for the operation of an interferometer are well described in the literature \citep[e.g.,][]{tho91}.  Briefly, an interferometer measures the time-averaged correlation 
of the complex electric fields measured at pairs of antennas  pointed at 
a particular object.  These correlations, or ``visibilities'' provide a 
measure of the spectrum of the sky brightness distribution at different 
spatial frequencies.  These frequencies are essentially the difference 
between the position vectors of the two antennas normalized by the 
wavelength of the observation in a coordinate system based on the position of the 
object in the sky.  These spatial frequencies are commonly referred to as 
$u$, $v$, and $w$, and the coordinate system used is defined such that $u$ is 
the spatial frequency in the east-west direction, $v$ corresponds to the 
north-south direction, and $w$ gives the spatial frequency along the line of 
sight to the object.  Thus, as the object moves through the sky, the $u$,$v$,$w$ 
coordinates of a pair of antennas, or ``baseline,'' changes.  The 
visibility measured for a specific baseline at a particular time is given by
\begin{equation}
V_{\nu}(u,v,w) = \int \int I_{\nu}(l,m) e^{-2 \pi i (ul+vm+nw)} d \Omega
\end{equation}
where $\Omega$ denotes solid angle, $\nu$ the frequency of the observed signal, 
and $I$ is the intensity on the sky at a 
position given by the direction cosines $l$, $m$, and $n= \sqrt{1-l^2-m^2}$ which 
are measured relative to the position of the observed object on the sky.\par
Generally, interferometers are ``fringe-stopped,'' that is, the measured visibilities are 
multiplied by a factor of $\mbox{exp}(-2 \pi i w)$ so that visibilities from a 
source at the center of the field of view will have a phase of zero (i.e., the 
source will not produce fringes).  This is mainly done because with fringe-stopping and for small fields of view 
(i.e., $n \simeq 1$), equation (1) becomes a simple two-dimensional Fourier transform such that 
the observed visibilities, measured as a function of $u$ and $v$, may be converted to 
maps of intensity on the sky using standard numerical methods.\par
As signals from astronomical sources pass through the ionosphere, a phase term is added 
given by
\begin{eqnarray}
\phi = \frac{e^2}{c m_e \nu} \int N_e(x) dx \\
\mbox{or, } \phi = 84.36 \left ( \frac{\nu}{\mbox{\scriptsize 100 MHz}} \right )^{-1} \left ( \frac{\mbox{\scriptsize TEC}}{\mbox{\scriptsize 1 TECU}} \right ) \mbox{ radians}
\end{eqnarray}
where $N_e$ is the electron density and $x$ denotes the path-length through the 
ionosphere.\par
Thus, the phase of the observed visibility for a baseline is altered by 
the difference between the ionospheric phase terms observed by the two antennas, which 
is proportional to the difference in the total electron content (TEC) observed along 
the lines of sight from the two antennas to the observed object.  To first order, if these 
phase terms are not removed, then during the Fourier inversion process involved in making an image of the sky, the ionospheric phase terms have the effect of changing the apparent positions of objects 
in the image plane.  When higher order ionospheric effects begin to dominate, 
objects begin to appear distorted in the image plane, and in extreme cases, almost disappear.\par
Therefore, to make an image, one must remove these phase terms through a calibration process 
which estimates the required phase corrections.  Typically, after initial calibrations for 
instrumental effects are performed, one usually uses some form of a procedure referred 
to as ``self-calibration'' \citep{cor99}.  This involves dividing the visibilities by an 
assumed sky model for the observed field of view which removes any contribution by the sky brightness distribution to the observed visibility phases.  Following this, a linear fit is used to determine the complex gain for 
each antenna.  Since the interferometer only provides phase differences between antenna pairs, 
the absolute phase of the complex gain for each antenna cannot be determined by this fitting 
process.  The general practice is to choose one reference antenna for the 
interferometer and to set the phase for its complex gain to zero.\par
Since for $N$ antennas, there are $N(N-1)/2$ baselines, this is generally 
an over-determined problem and can be done over relatively short time periods, depending 
on the brightness of the source.  One may also use this calibration to make an image, 
deconvolve the image to produce a better sky model, and then repeat the process until it 
converges.  This has been shown to be a rather robust procedure for determining the complex 
antenna gains \citep[see][and references therein]{cor99}.\par
According to equations (2) and (3), the phase corrections obtained from the determined antenna 
gains are essentially measurements of the difference between the TEC along the lines of sight of 
a particular antenna and that of the reference antenna.  These equations also demonstrate that 
the effect of the ionosphere will be more substantial for VHF observations.  Given the size 
of available astronomical VHF interferometers (baselines ranging from $<1$ km to as large as 
$\sim\! 30$ km), the robustness of the self-calibration procedure on relatively small time scales 
(typically $\sim\! 1$ minute, but as small as a few seconds for extremely bright sources), and 
the sensitivity of such interferometers to relatively small TEC variations (fluctuations in 
differential TEC of as small as 0.001 TECU), these instruments are capable of studying TEC 
fluctuations on substantially finer scales than many others available.  Subsequently, previous 
work has been done using astronomically motivated observations \citep[e.g.,][]{coh09,int09} as well as 
observations geared toward studying the ionosphere \citep[e.g.,][]{jac92a,jac92b} to 
explore the phenomenology of the ionosphere on these fine scales.\par
Much of this work has been performed with the VLA (latitude$= 34^{\circ} \: 04' \: 43.497''$ N and longitude$=107^{\circ} \: 37' \: 05.819''$ W).  This is in part due to the fact that the 
VLA is relatively well suited to the study of the ionosphere because its 27 antennas are distributed 
in a ``Y''-shape 
which allows it to probe structures along three different directions.  Arrays such as the WSRT and 
ATCA are so-called ``east-west'' arrays because the antennas are aligned along a single east-west 
axis and can therefore only observe ionospheric fluctuations along one dimension.  The VLA is 
also unique in that the antennas are moved from time to time among four different configurations, 
referred to A, B, C, and D.  The configurations go from larger and more spread out to smaller and more 
compact.  At its largest in the A configuration, each of the array's three arms is about 20 km long 
with the shortest antenna separations being about 1.5 km.  In the most compact configuration, D, the 
arms are at most 0.6 km long and the shortest spacings are about 0.04 km.  This allows the VLA the 
ability to study the ionosphere over a wider range of physical scales than other similar 
interferometers.\par
Finally, the VLA had a somewhat unique VHF system in place that allowed 
observations to be made simultaneously at 74 MHz and 327 MHz using a pair of dipole antennas 
(one for each band) mounted near the prime focus of each antenna.  Recently, the VLA electronics 
and receivers were upgraded to establish the new Expanded VLA (EVLA) which does not include 
the old VLA VHF system.  However, a new and improved VHF system is now being developed and will 
be available in the near future.\par
Past work using the VLA and other similar instruments 
has led to interesting results.  These include discoveries such as the new class of magnetic-eastward-directed 
(MED) waves, predominantly found at night, discovered by \citet{jac92b} as well 
as larger statistical studies such as the measurements by \citet{coh09} of the 
dependence of differential ionospheric refraction on relatively large angular scales ($>10^{\circ}$) using data from a 74 MHz 
all-sky survey which showed large dependences on time of day.  However, there is much more to 
be learned from these types of data, especially at smaller time and amplitude scales.  
Therefore, we are embarking on a program utilizing 
the VLA data archives (https://archive.nrao.edu) that seeks to push this type of 
analysis to even finer scales.  We will use previously unexplored data sets of the brightest 
VHF objects.  We will apply new techniques for calibration and the mitigation of radio frequency 
interference (RFI) to other data sets to significantly improve their sensitivity to small amplitude 
TEC fluctuations as well as fluctuations occurring on smaller time and spatial scales than could be 
explored previously.\par
Here, we describe the first step in this program, a thorough evaluation of 
the ionospheric information contained within a single, relatively long VHF observation of one of 
the brightest radio sources in the sky with the VLA.  This data set provides the opportunity to 
develop and establish techniques for processing, analyzing, and interpreting similar data in future 
segments of our program.  This paper focusses on the data selection and 
calibration as well as the post-processing done on the phase information extracted from this 
exemplar data set to measure TEC gradients with the array.  In a companion paper \citep{hel11}, we 
detail new techniques for spectral analysis of these data.

\section{Data Acquisition and Processing}
\subsection{Phase Correction Determination}
In the study presented in this paper, we have sought to explore the 
kinds of ionospheric phenomena capable of being observed with the 
VLA 
at the lowest fluctuation levels and the smallest times scales possible.  
To ensure that we were able to determine ionospheric phase 
terms to the the level of accuracy needed at the smallest available time 
intervals, including the resolution of any $2\pi$ ambiguities, we chose to 
use a data set focused on Cygnus A (or ``Cyg A''; also known as 3C405).  
With a total intensity of more than 17000 Jy ($=1.7 \times 10^{-22}$ 
W m$^{-2}$ Hz$^{-1}$) at 74 MHz, Cyg A is one of the two brightest sources in the sky at frequencies below 100 MHz.  Its brightness 
distribution on the sky is also known relatively well, implying that 
it can be used within the self-calibration procedure described briefly in \S 1 to solve for the phase corrections (both instrumental 
and ionospheric) for each antenna over relatively short time intervals.\par
\begin{figure}
\noindent\includegraphics[width=6in]{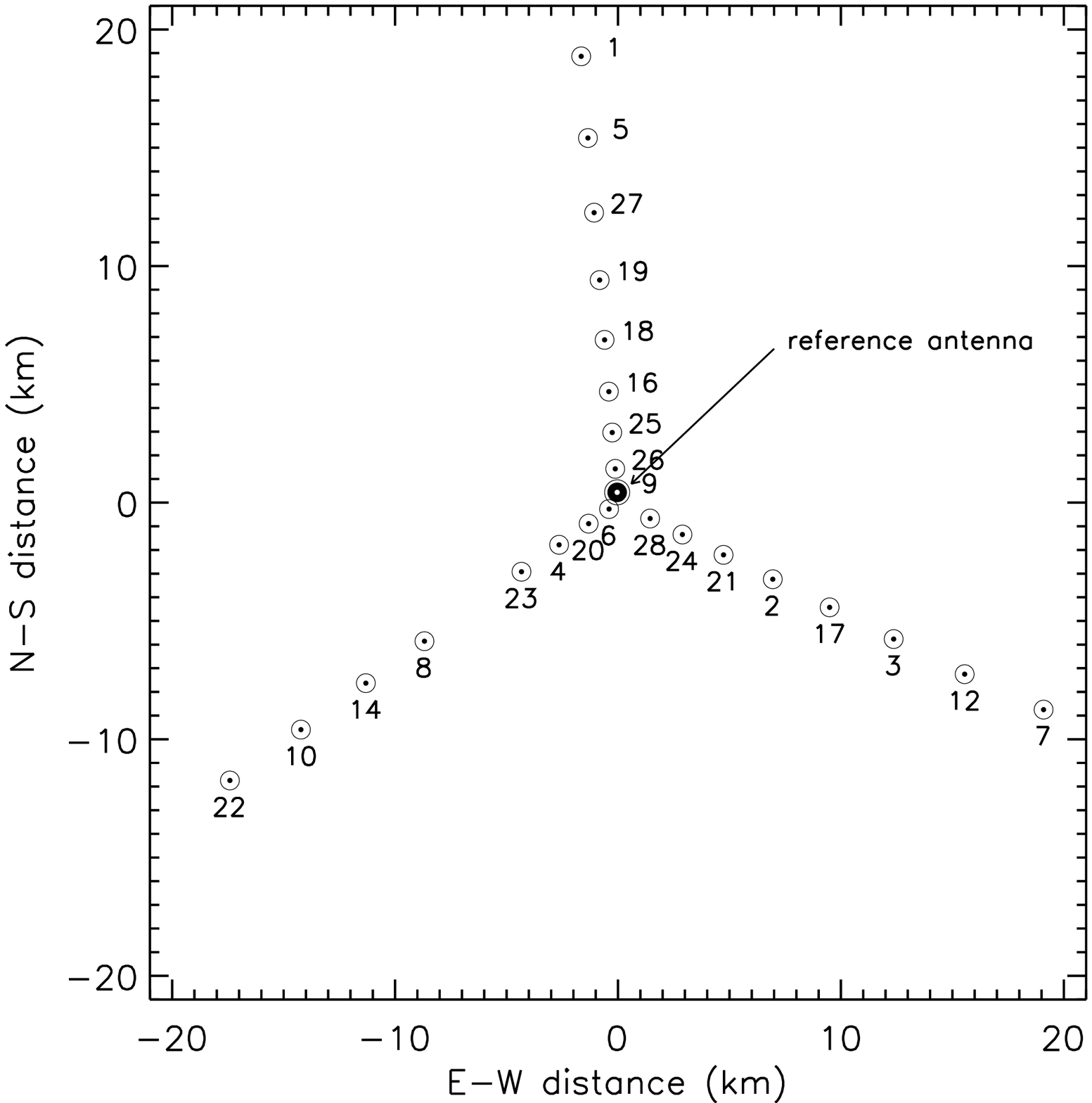}
\caption{The layout of the VLA antennas during the observations of Cyg A.  The reference 
antenna (see \S 2) is highlighted in black.}
\label{layout}
\end{figure}

The data set we have selected consists of simultaneous 74 MHz and 327 MHz 
dual polarization observations of Cyg A with the VLA over a period 
of more than 12 hours on 12-13 August, 2003 (VLA program number AK570).  During this period, there was a moderate amount of geomagnetic activity ($K_p$ index $\approx \! 2$--4) and low solar activity ($F10.7\!=\!123$ SFU; 1 SFU$\!=\!10^{-22}$ W m$^2$ Hz$^{-1}$).  
For this data, the VLA was in the A configuration with the VHF dipole system 
available on all but three antennas (antenna numbers 11, 13, and 15; see \S 1 and Fig.\ \ref{layout}).  The exact 
layout of the antennas is shown in Fig.\ \ref{layout}.  The data set contains a 35 
minute scan (i.e., a contiguous block of observing time), and a longer, nearly 
13 hour scan which, taken together, cover 
a range of local time from roughly $17^{\mbox{\scriptsize h}}00^{\mbox{\scriptsize m}}$ 
on 12 August to $06^{\mbox{\scriptsize h}}30^{\mbox{\scriptsize m}}$ the next 
morning.  For the 74 MHz band, the total bandwidth was 1.5 MHz; it was 
3 MHz for the 327 MHz band.  For both bands, the time-averaged 
visibilities were measured over intervals of 6.67 seconds simultaneously 
in both RR and LL polarizations.\par
The processing of VHF interferometric observations from the 
VLA or other similar interferometers is, in the average case, a lengthy 
and difficult process.  One usually must deal with significant sources 
of RFI, which typically appear much stronger 
on shorter baselines.  For the VLA, the field of view at 74 MHz is more than 
ten degrees in diameter, and the ionospheric conditions may vary significantly 
from one part of the field of view to another, which requires special 
calibration techniques.  One must also typically produce an image for the 
entire field of view, regardless of which object(s) one is interested 
in because the removal of sidelobes (i.e., secondary peaks of the impulse response in the image plane) produced by other objects is crucial 
to minimizing noise in the final image.  Fortunately, the unusually high 
intensity of Cyg A makes it substantially brighter than any source of RFI 
or any other object in the field of view.  This makes the calibration 
process much simpler and more straight forward.\par
The calibration of the data was performed using standard tasks within the 
Astronomical Image Processing System \citep[AIPS;][http://www.aips.nrao.edu]{bri94}.  
The first step in the calibration was to determine the bandpass response 
of each band (AIPS task \texttt{BPASS}).  This was done within one minute time 
intervals by dividing the data by the visibilities measure within a few 
central channels where the response typically peaks.  The relative amplitude 
and phase responses across each bandpass were then measured and interpolated 
onto the full set of time steps.  After trimming the first and last few channels 
from each band where the response drops substantially, the data were corrected for 
the bandpass responses.\par
Following this, at 6.67 second intervals (i.e., the 
shortest possible for these data), the data for each band and polarization 
were used with model visibilities computed using images of Cyg A 
presented by \citet{laz06} 
(currently publicly available at http://lwa.nrl.navy.mil/tutorial/) to 
compute the phase corrections for each antenna at each time step.  This was 
done with the AIPS task \texttt{CALIB} which does a series of consistency and 
sanity checks as it determines the solutions needed to fit the measured 
visibilities to the model and flags antennas and intervals that appear 
spurious or of poor quality.  We relaxed some of the criteria for these checks, namely 
the minimum number of antennas (we used four) required and the minimum signal to noise 
ratio (we used three) because of the relatively short time intervals used.  As noted above, 
an interferometer like the VLA can only measure relative phases, implying 
that for this calibration to work, a reference antenna must be chosen whose 
phase is arbitrary and is subsequently set to zero.  Typically, the antenna 
closest to the center of the VLA is not used because the effects of RFI tend 
to be the worst for this antenna.  However, as noted above, Cyg A is bright 
enough that this is not a consideration for our data set, and we therefore 
used this antenna as our reference (antenna 9; see Fig.\ \ref{layout}).

\subsection{Processing of Phase Corrections}
Following the determination of phase corrections using standard AIPS 
routines, several steps were taken to extract ionospheric information from 
the phase data.  This was done using ad hoc, python-based software.\par
The phase corrections measured by the calibration process contain 
contributions from several effects, the ionosphere being the largest, 
especially at 74 MHz.  In particular, the phase difference between two antennas within a single baseline is given by
\begin{equation}
\Delta \phi = \Delta \phi_{ion} + \Delta \phi_{inst} + \Delta \phi_{sour} + \Delta \phi_{amb}
\end{equation}
where $\Delta \phi_{ion}$ denotes the difference in the ionospheric phases along the lines-of-sight of the two antennas given by equation (3), $\Delta \phi_{instr}$ represents difference in the instrumental effects of the two antennas, $\Delta \phi_{sour}$ is the contribution to the phase difference from the structure of the observed source, and $\Delta \phi_{amb}$ is the contribution from $2\pi$ ambiguities.  We can remove $\Delta \phi_{sour}$ by dividing the observed visibilities by a model of Cyg A and $\Delta \phi_{amb}$ by having short enough time sampling to ``unwrap'' the phases (see below).  
The dual 74 and 327 MHz observing mode is very useful 
for removing instrumental effects since the ionospheric phase simply scales 
with wavelength [see equations (2) and (3)], whereas instrumental effects do not 
necessarily.  The instrumental components of the phase corrections include 
errors in the delays added to individual antenna signals used to fringe-stop the visibilities (see \S 1) and offsets between the antenna pointing and the actual source position.  These two effects and $2\pi$ ambiguities were dealt with in three separate steps.\par
First, the AIPS task \texttt{CALIB} flags antennas that are spurious at each time step 
so that the phase corrections are ``missing'' for some antennas at a relatively 
small fraction ($\sim 2\mbox{\%}$) of time steps.  Generally, a small amount of missing data is not problematic.  However, to facilitate the use of fast Fourier transforms (FFTs) within our spectral analysis of the data presented in a companion paper which work best with evenly sampled data, these missing time steps were filled in.  We have done this by linearly interpolating the real and imaginary parts of the complex 
antenna gains, $g_{A}$, computed by \texttt{CALIB} for all antennas onto a common 
time-step grid consisting of 7299 steps spaced by 6.67 seconds, and then 
recomputing the phase corrections \{$=\mbox{tan}^{-1}[\mbox{Im}(g_A)/\mbox{Re}
(g_A)]$\}.\par
\begin{figure}
\noindent\includegraphics[width=6in]{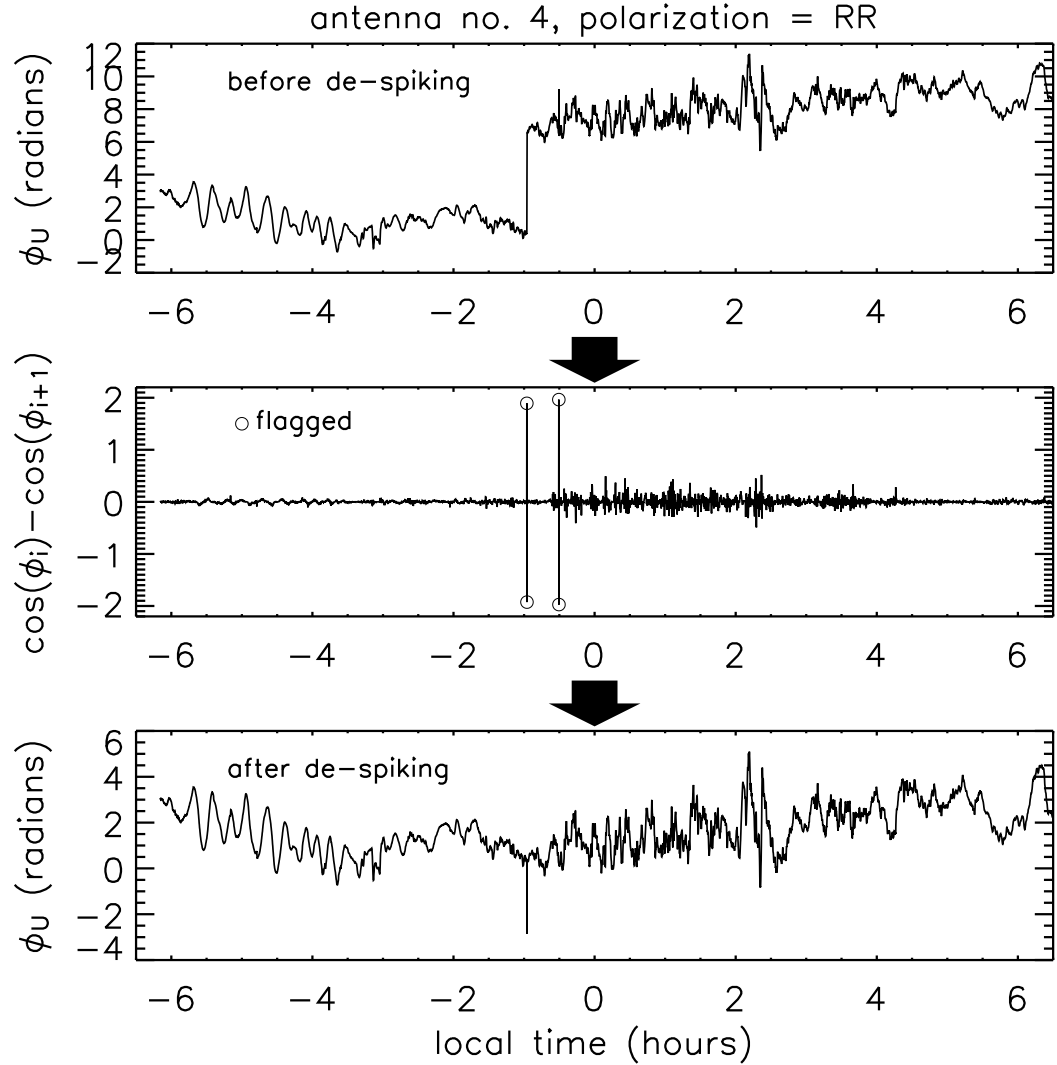}
\caption{Upper panel: the unwrapped phase, $\phi_{\mbox{\scriptsize U}}$ for antenna 4, 
RR polarization, as a function of time (relative to midnight, 13 August) 
when no attempt was made to ``de-spike'' the phase 
data (see \S 2.2).  Middle panel: the difference between 
the cosine of the wrapped phase at a 
time step i, $\mbox{cos}{\phi_{\mbox{\scriptsize i}}}$, and the next time step as a 
function of time used to find spikes in the phase data.  Flagged spikes are highlighted 
with circles.  Lower panel: the phase as a function of time, unwrapped with the flagged 
spikes excluded.}
\label{despike}
\end{figure}

Next, the time sampling of the data (6.67 seconds) was sufficiently short that 
the phase, $\phi$, as a function of time for each antenna could be ``unwrapped'' 
in the conventional way, i.e., by correcting phase jumps of more than $\pi$ 
radians by adding or subtracting $2 \pi$.  There was one caveat to this process, 
however.  For each antenna, there were a few (ranging from zero to five) times 
steps where, for one reason or another, the phase correction was either 
spurious or represented a real and very short jump in the instrumental phase, appearing as sharp spikes in the unwrapped phases.  Since there were 
a total of 7299 time steps, a few short jumps in phase would not be an issue except for 
their effect on the unwrapping process.  Any of these spikes can cause an artificial 
large phase jump if it is included in the unwrapping process, which we have 
illustrated in the upper panel of Fig.\ \ref{despike}.\par
To combat this, we 
wrote a simple algorithm that computes the difference between $\mbox{cos}(\phi)$ at 
each time step and the value for the next time step, where $\phi$ is the wrapped 
phase.  Any time step where the absolute value of this difference was more than ten 
times the standard deviation among all time steps for a give antenna was flagged and 
not included in the unwrapping process.  These empty time steps were then filled by 
linearly interpolating the unwrapped phase data for the un-flagged time steps.  It should be noted that these spikes only occupy 1--2 time steps (6.67--13.34 seconds) and that the surrounding data are otherwise well-behaved, making interpolation a reasonable and straightforward solution to eliminating instances of such spikes.  An example of 
how the data were flagged is illustrated in the middle panel of Fig.\ \ref{despike} 
while the resulting unwrapped phases are shown in the lower panel.  From this result, 
one can see that the spikes are not always completely removed from the data, but the 
goal of eliminating their effect on the unwrapping process has been achieved.\par
\begin{figure}
\noindent\includegraphics[width=6in]{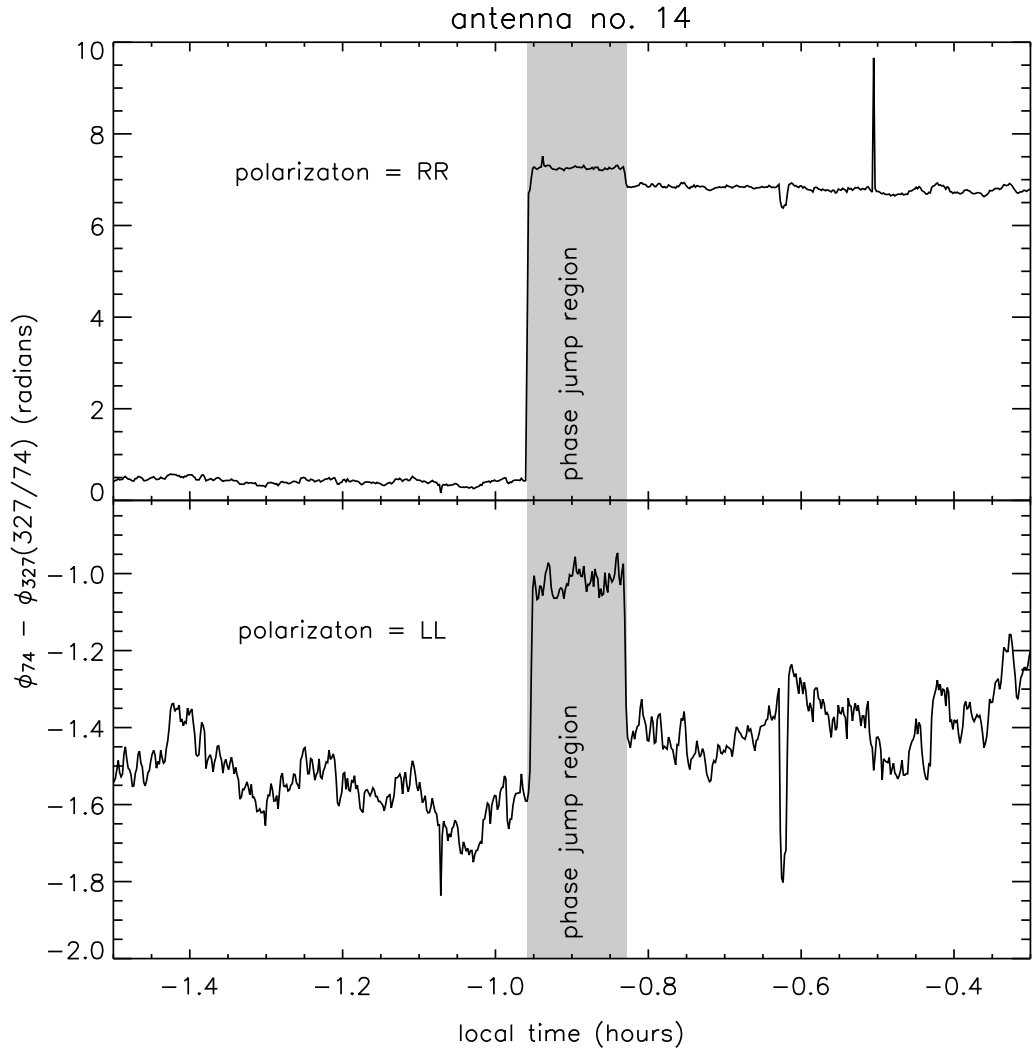}
\caption{As a function of time, the difference between the unwrapped 74 MHz phase, 
$\phi_{74}$, and the 327 MHz phase, $\phi_{327}$, with the 327 MHz phase scaled such 
that the ionospheric components of the two phases canceled out for the RR (upper) 
and LL (lower) polarizations for antenna 14.  The region where the instrumental 
phases have changed relatively abruptly (i.e, the ``phase jump'' region first referred 
to in \S 2.2) is shaded in gray.}
\label{pjump}
\end{figure}

A number of instrumental effects can contribute to the phase corrections, including 
errors in the fringe-stopping process (see \S 1) and offsets between the position of the source and that of the observed field center (i.e., pointing errors).  These effects are generally stable 
in time, changing insignificantly over periods of days \citep[see, e.g.,][]{coh07}.  
However, for 
the VLA, the instrumental phase is known to occasionally have short jumps that one 
must be wary of.  We have found such a jump in our data occurring at a local time of 
about $23^{\mbox{\scriptsize h}}06^{\mbox{\scriptsize m}}$ on 12 August.  The jump is most obvious when one plots the difference 
between the 74 MHz phase and the 327 MHz phase scaled by a factor of (327/74) so that 
the ionospheric phases cancel out (note, this only gives us the scaled difference between the 327 and 74 MHz instrumental phases, not the instrumental phases themselves).  We have plotted this difference as a function of 
time for antenna 14 for both polarizations in Fig.\ \ref{pjump} to show the location 
of what we will refer to as the ``phase jump region.''  While being fairly subtle in 
the LL polarization, it is quite obvious in the RR polarization.  The jump lasted 
about 6 minutes and can be seen in the data for several antennas.  To deal with this, 
we have treated the phase jump region, as well as the time periods before and after 
it, as separate scans, assuming that each scan has its 
own instrumental phase.  This basically resulted in us treating the data as if it 
contained four scans instead of two.\par
To remove instrumental effects, we have used a kind of continuum 
subtraction process. Within this process, we have treated ionospheric fluctuations as features 
superimposed on a smooth continuum consisting of the instrumental phases, which vary relatively slowly with time as well as any slowly varying component of the ionospheric phase.  With the current data, we 
unfortunately do not have the means to separate the slowly varying component of the ionospheric phase from the instrumental effects and can therefore only measure 
fluctuations in TEC on relatively short ($<1$ hour; see below) time scales.  In the future, with instruments with 
larger bandwidths, it will be possible to use the wavelength dependence of the 
phase corrections to separate these effects since the instrumental phases are $\propto \nu$ and the ionospheric phase is 
$\propto \nu^{-1}$.\par
We have chosen to perform our continuum subtraction process for each antenna, band, 
and polarization 
by smoothing the unwrapped phases with a one-hour-wide boxcar which appeared to 
preserve any apparent fluctuations while giving a good representation of the continuum.  
For the first scan and the phase jump region, we simply subtracted a single mean 
value from all the phases since they are each shorter than an hour.  We did the 
same for the first and last hour of each of the remaining two scans so that the 
same filter width would be used for all times steps.\par
Following this, we found that the position offset component of the instrumental phases presented a problem for 
this process near the edges of each scan.  This is because, depending on the 
antenna, these phases can vary significantly over one hour, especially at 327 MHz.  
From the fringe-stopped (see \S 1) version of equation (1), one can see that a 
position offset in the direction cosines $l$ and $m$ of $\Delta l$ and $\Delta m$ 
will produce an additional phase of $-2 \pi (u \Delta l + v \Delta m)$.  Since 
$u$ and $v$ are normalized by the observed wavelength, any such phase will be 4.4 
times larger for the 327 MHz band.  In addition, the offsets can be different for each band and this difference can vary with baseline.  This is due to a number of factors including the fact that different model images were used for the bands which may not be exactly aligned and that Cyg A has a significant amount of resolved structure which larger baselines are more sensitive to, especially at 327 MHz where the angular resolution in the image plane is 4.4 times better than that at 74 MHz.\par
\begin{figure}
\noindent\includegraphics[width=6in]{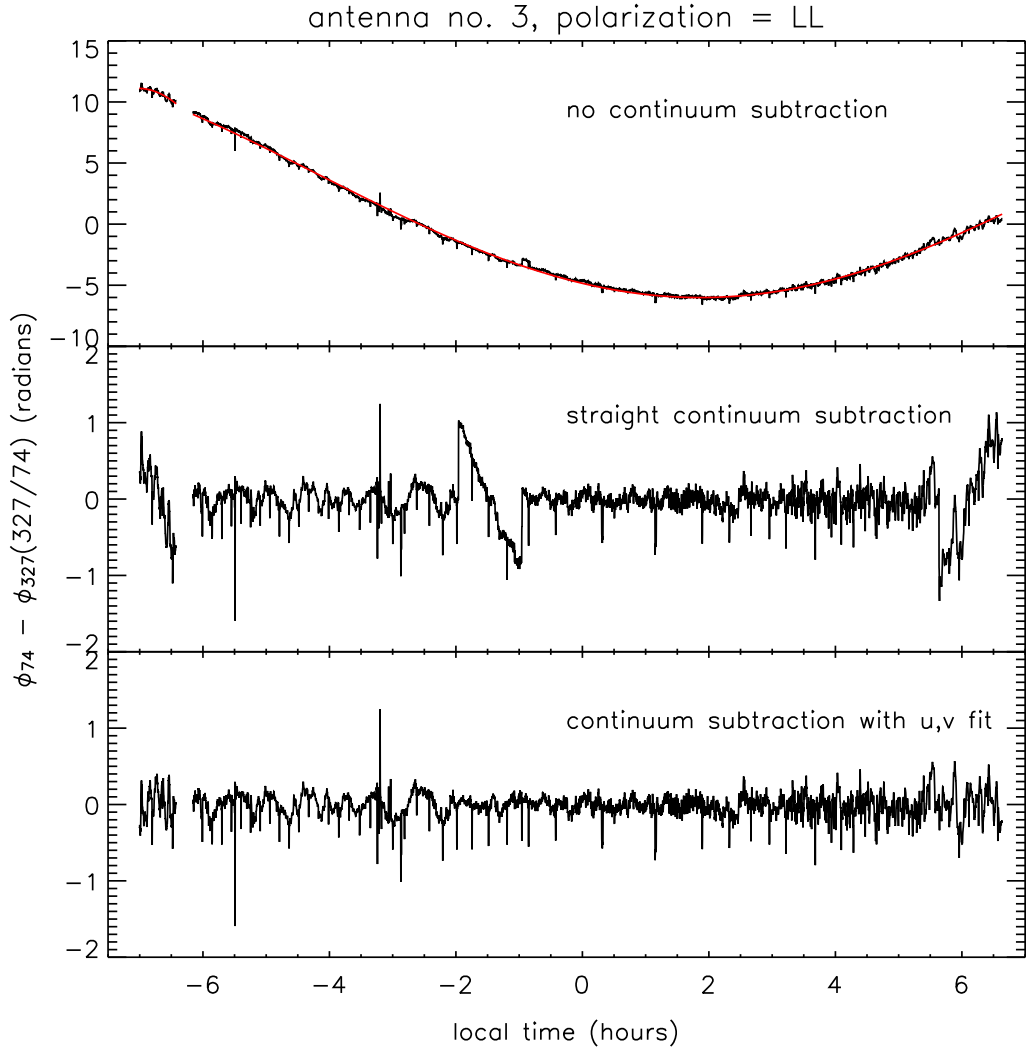}
\caption{Upper: As a function of time, the difference between the unwrapped 74 MHz phase, 
$\phi_{74}$, and the 327 MHz phase, $\phi_{327}$, with the 327 MHz phase scaled such 
that the ionospheric components of the two phases canceled out for antenna 3, LL 
polarization.  A fit to this data of a linear combination of the spatial frequencies 
$u$ and $v$ [see equation (1)] which assumes 
a single position offset (see \S 2.2) is also plotted in red.  Middle:  The same as 
in the upper panel, but for the phases with the smoothed (with a one-hour wide box) 
phases subtracted from each band.  Lower:  The same as the middle panel, but with a 
fit of a linear combination of $u$ and $v$ used in the continuum subtraction of the 
327 MHz phases at the edges of each scan (see the discussion in \S 2.2).}
\label{edge}
\end{figure}

We have demonstrated this in the upper panel of Fig.\ \ref{edge} where we have plotted 
$\phi_{74} - \phi_{327} (327/74)$ for antenna 3, LL polarization as a function of 
time.  With $\phi_{327}$ scaled by (327/74), the ionospheric phases are removed and 
all that is left is the difference between the instrumental 
phases for the two bands.  The data follow a smooth curve which is inconsistent with 
the known behavior of VLA instrumental phases, especially at 74 MHz.  Furthermore, the curve that the 
data follows is easily fit by a linear combination of the un-normalized versions 
of the $u$ and $v$ coordinates (see the red curve in the upper panel of Fig.\ 
\ref{edge}).  A single baseline (antenna 3 with 
the reference antenna) will sweep out an ellipse in the $u$,$v$-plane because 
of the rotation of the earth \citep[see, e.g.,][]{tho99}.  Therefore, this is exactly 
what one would expect for a scenario where there is a single position offset 
for each of the two bands during each scan.\par
To show the effect of the time dependence of the position offset phase on 
our continuum determination process, we have plotted the continuum-subtracted 
version of $\phi_{74} - \phi_{327} (327/74)$ versus time in the middle panel of Fig.\ \ref{edge}.  
One can see from this plot that within the first scan and within the last hour of the 
second and fourth scans where a single mean continuum value was subtracted from each, 
the gradient of the position offset phases has introduced an artificial difference 
which increases/decreases with time.  Since the position offset phase is much larger 
for the 327 MHz band, we have introduced the following additional step for the 
continuum subtraction of the 327 MHz data.  Within the first scan, the phase jump 
region, and the first and last hours of the other two scans, we have fit a linear 
combination of the un-normalized versions of $u$ and $v$ to $\phi_{74} - \phi_{327} (327/74)$ 
separately for each time range.  Within each of these time periods, we used the 
mean values for $\phi_{74}$, $\phi_{372}$, and the $u$,$v$ fit to construct a 
time-variable continuum for the 327 MHz data.  The benefits of this are illustrated 
in the bottom panel of Fig.\ \ref{edge} where we have plotted the continuum-subtracted 
version of $\phi_{74} - \phi_{327} (327/74)$, this time, including the additional 
computation for the time-variable 327 MHz continuum with the first/last hour of each 
scan.  One can see that the roughly linear features seen in the middle panel of 
Fig.\ \ref{edge} have been removed and that the remaining difference between the 
74 MHz and 327 MHz continuum-subtracted phases is essentially noise.\par
Following the application of the corrections detailed above, we used equation (3) 
to convert the continuum subtracted phases for each antenna, band, and polarization 
to values of differential TEC, or $\delta \mbox{TEC}$.  Then, at each time step and 
antenna, we computed the median $\delta \mbox{TEC}$ among the four values (i.e., two 
bands and two polarizations) as well as the median absolute deviation (MAD) as an 
estimate of the uncertainty in the median.  To increase the reliability of the MAD 
computations, we included with each time step the four nearest time steps (using 
their own individual median values) for a total of 20 data points per time step.  
In both computations (median and MAD), the median was 
used to minimize the effects of any spurious data which remained.\par
\begin{figure}
\noindent\includegraphics[width=6in]{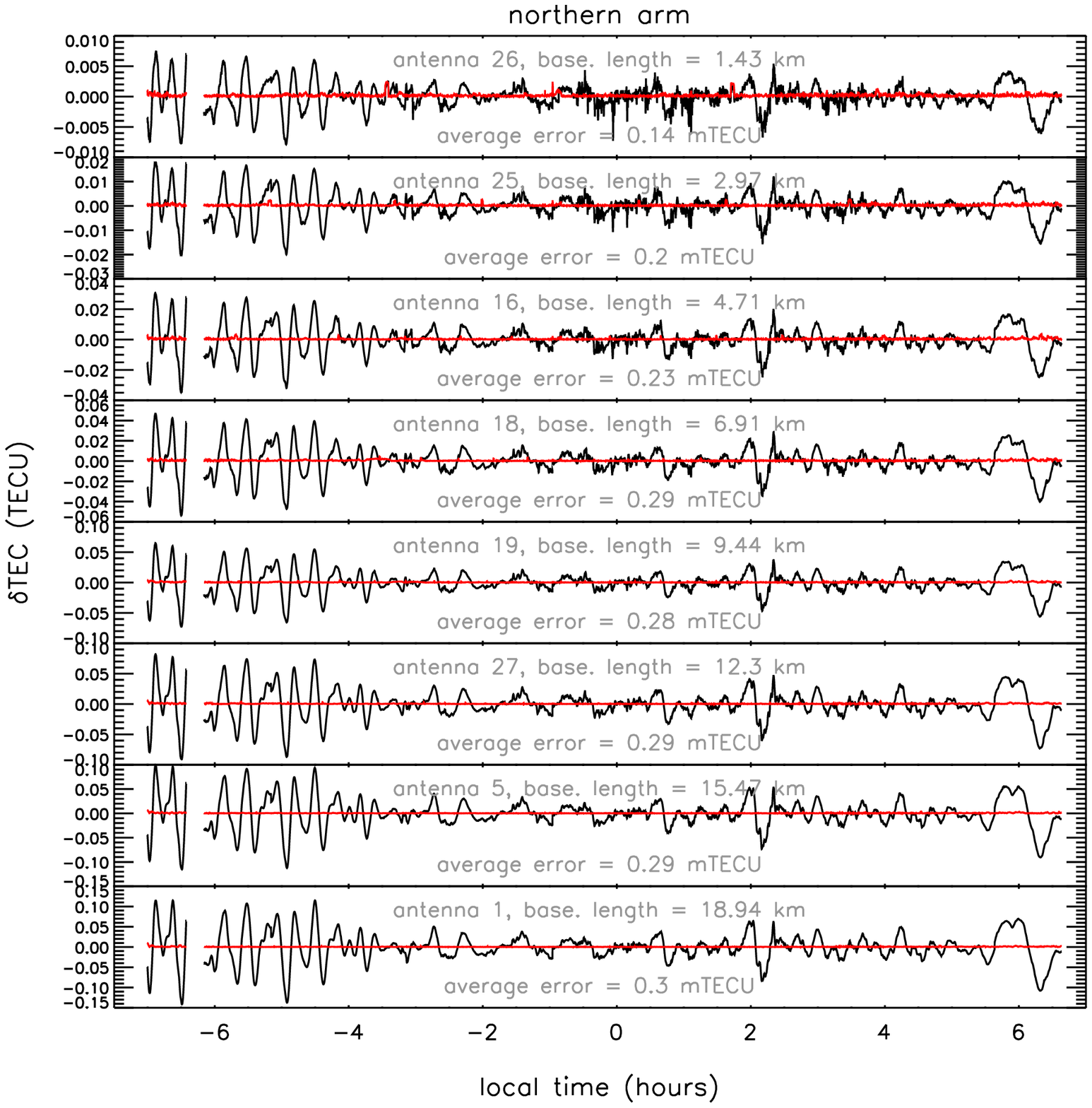}
\caption{For each antenna in the northern arm (see Fig.\ \ref{layout}) of the VLA, the 
the difference between the TEC fluctuation (i.e., above or below the mean TEC) along its 
line of sight and that measured along the reference antenna's line of sight, $\delta 
\mbox{TEC}$.  The estimated uncertainty (see \S 2.2) is plotted in red in each panel.}
\label{north}
\end{figure}
\begin{figure}
\noindent\includegraphics[width=6in]{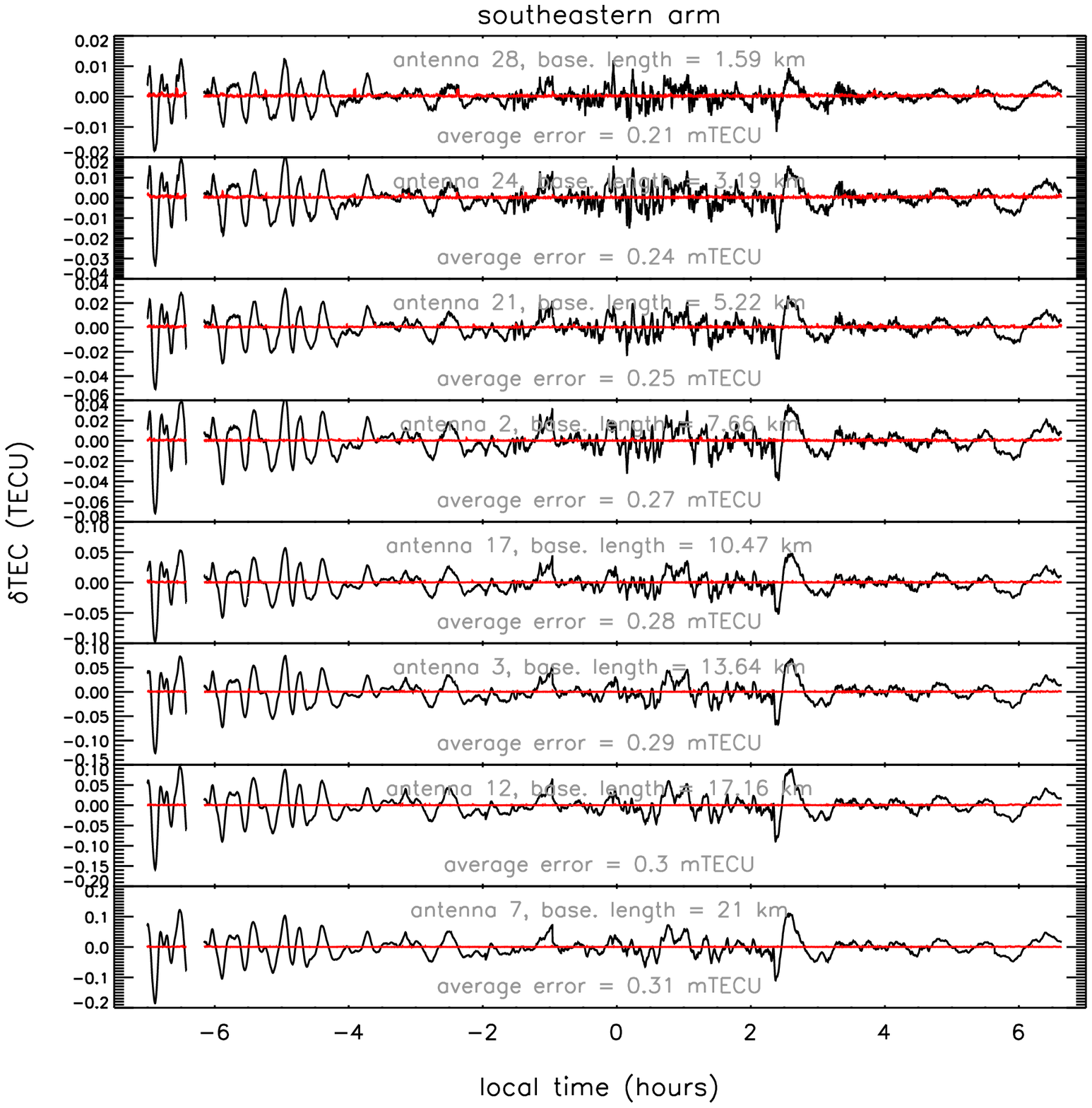}
\caption{For each antenna in the southeastern arm (see Fig.\ \ref{layout}) of the VLA, the 
the difference between the TEC fluctuation (i.e., above or below the mean TEC) along its 
line of sight and that measured along the reference antenna's line of sight, $\delta 
\mbox{TEC}$.  The estimated uncertainty (see \S 2.2) is plotted in red in each panel.}
\label{east}
\end{figure}
\begin{figure}
\noindent\includegraphics[width=6in]{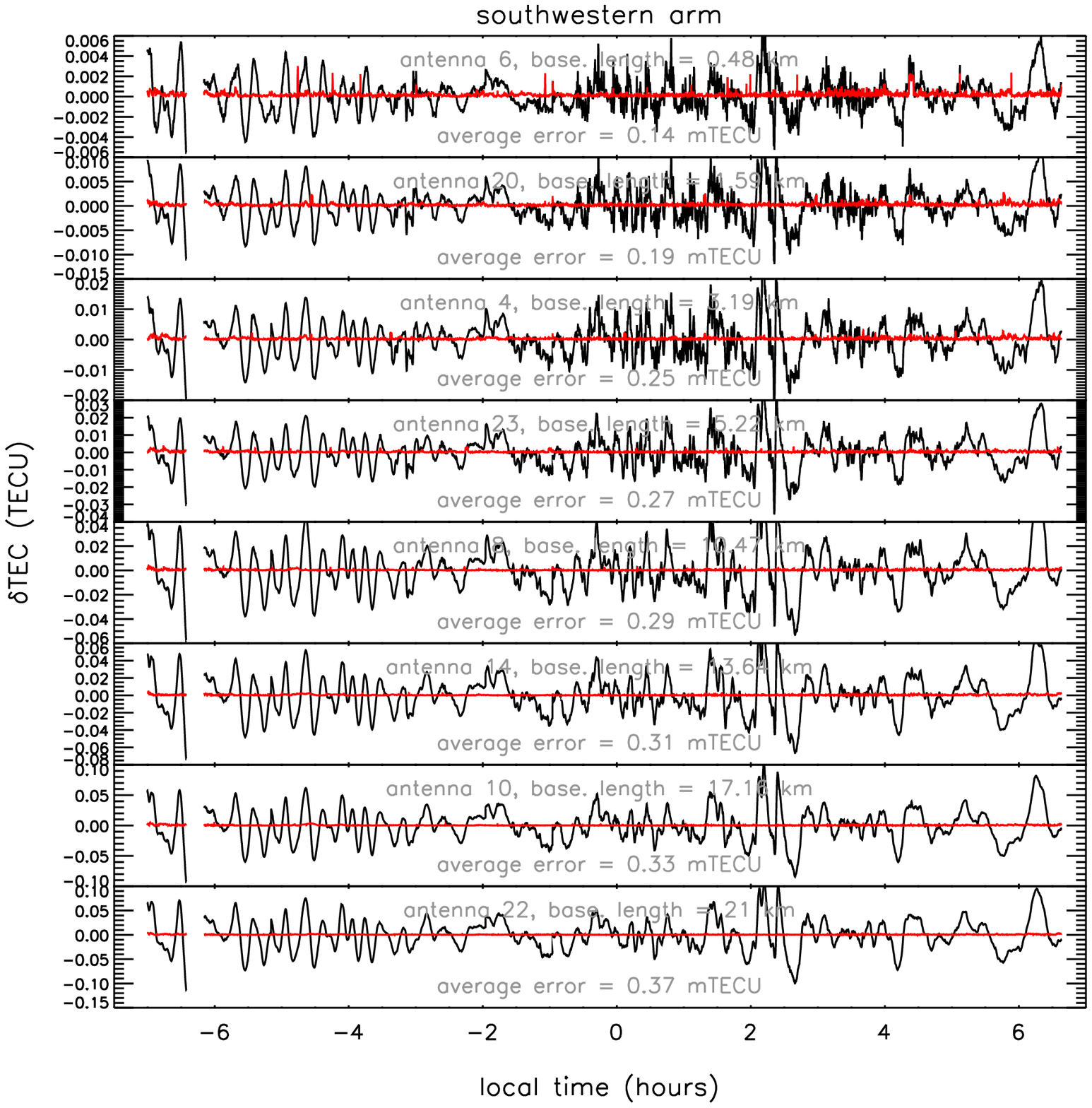}
\caption{For each antenna in the southwestern arm (see Fig.\ \ref{layout}) of the VLA, the 
the difference between the TEC fluctuation (i.e., above or below the mean TEC) along its 
line of sight and that measured along the reference antenna's line of sight, $\delta 
\mbox{TEC}$.  The estimated uncertainty (see \S 2.2) is plotted in red in each panel.}
\label{west}
\end{figure}

The resulting 
$\delta \mbox{TEC}$ values are plotted as functions of time for each antenna in the 
northern arm in Fig.\ \ref{north}, the southeastern arm in Fig.\ \ref{east}, and the southwestern arm 
in Fig.\ \ref{west} along with the MAD values to illustrate the relatively high 
precision to which $\delta \mbox{TEC}$ has been measured.  The typical 
$\delta \mbox{TEC}$ uncertainty, represented by the MAD computations, is about 
$3 \times 10^{-4}$ TECU, demonstrating the remarkable ability of the VLA to detect extremely 
small TEC fluctuations, even on time scales $<10$ seconds when an object as bright 
as Cyg A is used.

\section{Measuring TEC Gradients}
\subsection{General Approach}
With the fully reduced $\delta \mbox{TEC}$ data, including robust estimates of 
the uncertainties, we are in a position to explore a wide range of ionospheric phenomena.  
First, we note that since the VLA measures differential TEC values between 
antenna pairs, it is essentially only sensitive to changes in the TEC gradient.  
Given the geometry of the array (see Fig.\ \ref{layout}), we cannot numerically 
compute the TEC gradient at each antenna location from our data, and measuring the full TEC gradient requires a somewhat ad hoc approach.  Such measurements are crucial for any analysis of observed TEC fluctuations because without modification, the set of $\delta \mbox{TEC}$ time series can only be spectrally analyzed for specific assumed pattern models \citep[e.g., a single plane wave, see][]{jac92a}.\par
This is different from the normal mode of operation for radio interferometers in which standard techniques are used to invert and de-convolve sparsely sampled visibility data to make an image.  As equation (1) demonstrates, the observed visibilities are functions of the \emph{differential} antenna positions, $u$, $v$, and $w$.  For small fields of view, the contribution of the $w$ term is negligible if fringe-stopping is applied (see \S 1).  Thus, even for a Y-shaped array, reasonably good $u$,$v$-coverage can be obtained.  This is improved further by the rotation of the earth which causes each baseline to sweep out an ellipse in the $u$,$v$-plane \citep[see \S2.2 and][]{tho99}.  Using this fact to obtain better $u$,$v$ coverage is sometimes referred to as ``earth rotation synthesis.''\par
In contrast, the gradient of an arbitrary set of TEC fluctuations varies over the array as a function of the \emph{actual} antenna positions projected onto the ionosphere.  Improved spatial coverage can be obtained by exploiting the change in the apparent position of the observed source (i.e., rotation of the earth) and the movement of the fluctuations themselves, effectively converting temporal baselines into spatial ones.  However, since the TEC fluctuations presumably have a distribution of speeds and directions, this is not as straightforward as in earth rotation synthesis.  One must decompose the time series into temporal spectral modes and then analyze how the properties of each mode vary across the array to extract the size, speed, and direction of the dominant pattern(s) for that mode (such spectral techniques are detailed in a subsequent paper).  Therefore, we must still contend with data that has been sampled in a Y-pattern which cannot be inverted in a straightforward manner.  \par
We have consequently developed two ad hoc techniques designed to provide measurements of the TEC gradient time series over the full array and along each of the VLA arms.  Before implementing either technique, we first had to perform two 
basic geometric corrections to the data so that the measured TEC gradients 
would correspond to vertical TEC gradients as closely as possible.  First, to ensure 
that our characterization of the shape of the observed TEC surface is physically 
meaningful, we needed to project the antenna pattern displayed in Fig.\ \ref{layout} 
onto the locations where the lines of sight of the antennas pass through the 
ionosphere, or ``pierce-points.''  Second, we needed to compute the slant-to-vertical 
TEC corrections for the line of sight to Cyg A as its apparent position on the sky 
changed throughout the observation.  
For a plane parallel approximation, both of these 
corrections are relatively straightforward.  However, since Cyg A was as low as 
$12^{\circ}$ above the horizon during the observing run, a plane parallel 
approximation was far from valid at all time steps.  We have therefore computed 
the two required geometric corrections using a spherical model detailed 
in Appendix A.\par  Within this model,  
the ionosphere was approximated with a thin shell located at the height of the 
maximum electron density, or ``peak height'' \citep[see][]{lan88,ma03}.  We 
obtained estimates of the peak height as a function of time by using the International 
Reference Ionosphere \citep[IRI;][]{bil01} software, inputting the date and time of our observations 
and the latitude and longitude of the VLA.  We then re-determined the peak heights using 
the latitudes and longitudes of the pierce-points.  We found that additional 
iterations of this process only marginally changed the results and chose to use one 
iteration only.  The final peak heights used are plotted in the upper 
panel of Fig.\ \ref{zion} along with the projected separations from the array center for the 
farthest antennas of each arm (antennas 1, 7, and 22; see Fig.\ \ref{layout}) and the 
corresponding slant-TEC corrections.

\subsection{Polynomial Fits}
After applying the geometric corrections to the antenna positions and 
the $\delta \mbox{TEC}$ measurements, we sought to characterize the full two-dimensional TEC gradient observed by each antenna at each time step.  Rather than assume a particular dominant structure (e.g., a plane wave), we simply assumed that since the array is smaller than many transient ionospheric phenomena, the observed TEC surface at any time step could be approximated with a low-order Taylor series.  We examined many time steps and found 
that a second order, two-dimensional Taylor series adequately approximated the amount of curvature in the TEC surface detected by 
the VLA.  This Taylor series 
has the following form
\begin{equation}
\mbox{TEC} = p_{0}x + p_{1}y + p_{2}x^{2} + p_{3}y^{2} + p_{4}xy + p_{5}
\end{equation}
where $x$ and $y$ are the north-south and east-west antenna positions, respectively, 
projected onto the surface of the ionosphere at the estimated peak height.  
To maximize the amount of data used to constrain the parameters of each fit, we 
used the difference between $\delta \mbox{TEC}$ for each of the 300 unique antenna 
pairs at each time step.  Thus, the form of equation (5) actually fit to the data 
was
\begin{eqnarray}
\delta \mbox{TEC}_i - \delta \mbox{TEC}_j = p_0(x_i-x_j)+p_1(y_i-y_j) \nonumber \\
+p_2(x_i^2-x_j^2)+p_3(y_i^2-x_j^2)+p_4(x_iy_i-x_jy_j)
\end{eqnarray}
where the $i$ and $j$ subscripts denote the values for the $i^{\mbox{\scriptsize th}}$ 
and $j^{\mbox{\scriptsize th}}$ antennas, respectively.  We also utilized some 
standard sigma-clipping during the fitting process for each time step by computing 
the rms of the fit residuals, rejecting all antenna pairs with absolute 
residuals $>3 \, \mbox{rms}$, and repeating 50 times.  As many as about 10 and 
as few as zero were rejected for any given time step. We note that each time step 
was fit independently to preserve the presence of any small-scale spatial/temporal 
fluctuations.\par
\begin{figure}
\noindent\includegraphics[width=6in]{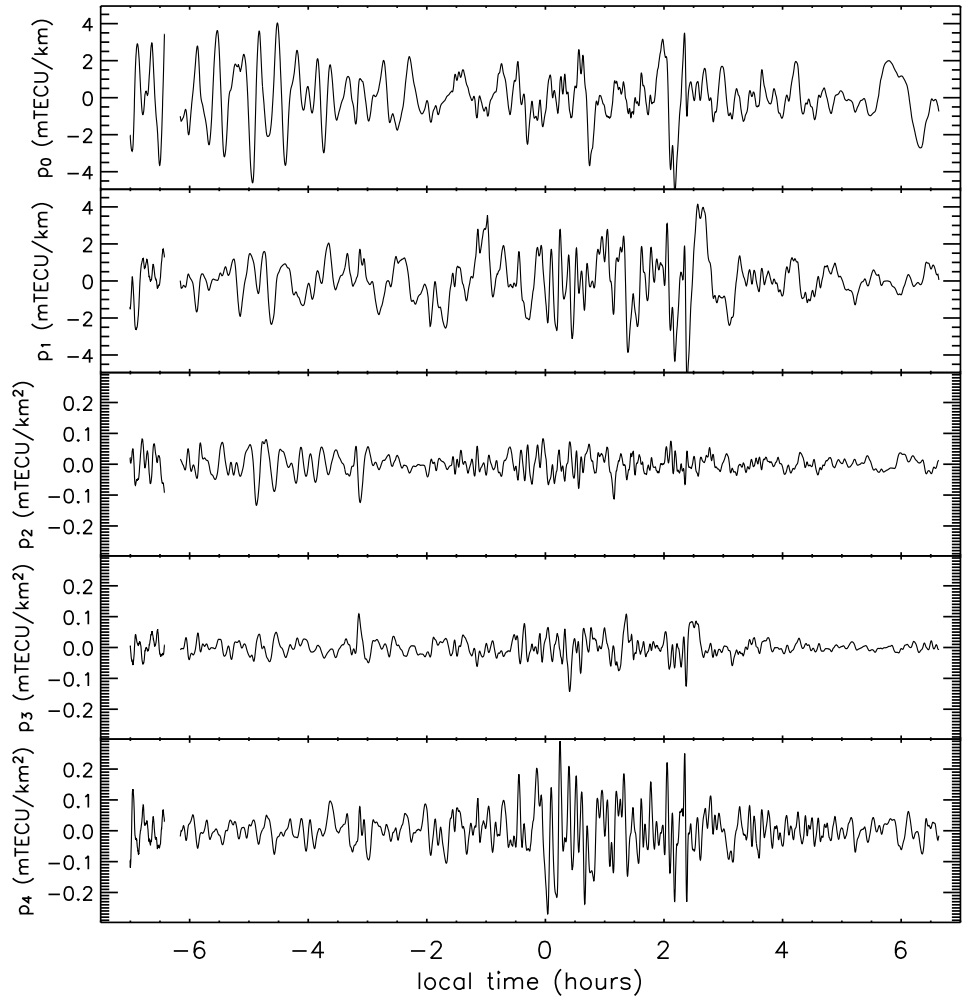}
\caption{The fitted values of the coefficients from equation (6) as a 
function of local time relative to midnight for 13 August, 2003.}
\label{pfit}
\end{figure}

The fitted coefficients as a function of time are plotted in Fig.\ \ref{pfit}.  
From these, one can see the same large amplitude and period 
fluctuations at the beginning of the 
observing run that are visible in the individual antenna data plotted in Fig.\ 
\ref{north}--\ref{west}.  Note that they are not quite as large here because of the 
applied slant-TEC correction discussed above.  Here, we can see that they are most 
visible in the $p_0$ coefficient which is the 
partial derivative of the TEC surface at the center of the array along the north-south 
direction.  With the 
plots in Fig.\ \ref{pfit}, one can also see the same thing beginning to happen 
near the end of the run toward dawn.  This is qualitatively consistent with the 
known behavior of medium-scale traveling ionospheric disturbances (MSTIDs) which 
are prevalent near sunrise and sunset \citep[e.g.,][]{her06}.  During the middle 
of the night, the second-order terms become more significant.

\subsection{Arm-based Approach}
While the polynomial-based measurements provide useful information about the variation of the full two-dimensional TEC gradient, they neglect the ability of the VLA in its A configuration to detect 
fluctuations on scales as small as a few kilometers.  In principle, one could do this by simply increasing the order of the polynomials used.  However, it is likely that the small-scale structures observed do not span the array.  This implies that such fits would not yield accurate representations of the full TEC gradient at each antenna, especially those near the ends of the arms (see Fig.\ \ref{layout}).  Therefore, instead of using higher order polynomial fits, we have opted for an alternative approach to make full use of the data.\par
\begin{figure}
\noindent\includegraphics[width=6in]{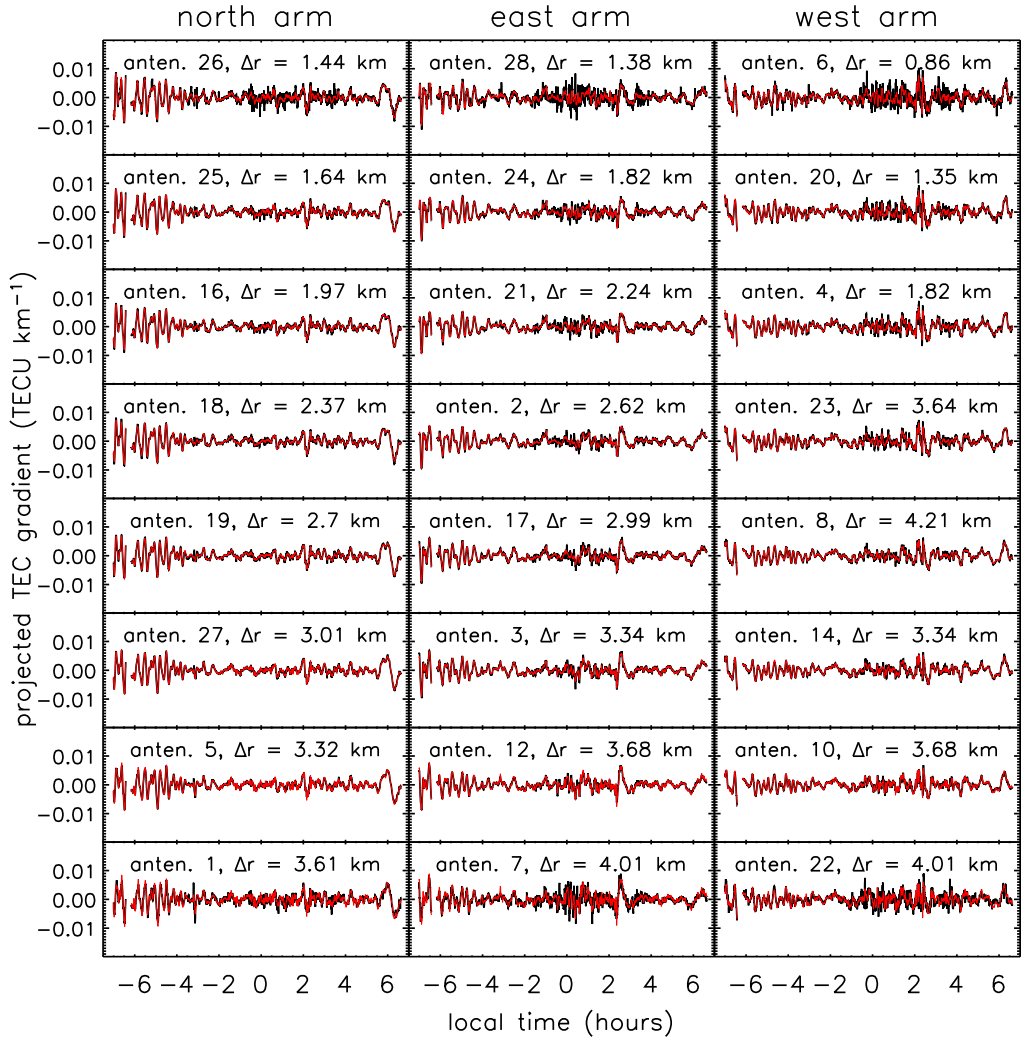}
\caption{The projection of the TEC gradient along each VLA arm (see Fig.\ \ref{layout}) at each antenna as a function of local time (black curves).  Plotted in red are the projected TEC gradients computed using the fitted polynomial coefficients plotted in Fig.\ \ref{pfit}.  In each panel, the antenna number and the mean separation among the nearest antennas ($\Delta \mbox{r}$) are printed.}
\label{tecgrad}
\end{figure}

This complementary method computes the projection of the TEC gradient (or, the spatial derivative of $\delta \mbox{TEC}$) along each VLA arm.  The projected gradient was computed at each time step separately for the antennas of each arm using simple three-point Lagrangian interpolation.  Given the typical $\delta \mbox{TEC}$ precision of $3 \times 10^{-4}$ TECU and the mean separation between antennas of 2.5 km, the precision of these projected TEC gradient measurements is typically about $2 \times 10^{-4}$ TECU km$^{-1}$.  The time series of the projected gradient at each antenna is plotted in Fig.\ \ref{tecgrad}.  We have also plotted in red the projected gradient computed using the polynomial coefficients plotted in Fig.\ \ref{pfit}.  One can see that for the larger amplitude, longer period disturbances, the polynomial fits largely recover the structure observed using the data for individual antennas.  However, during the middle of the night, there appears to be a significant amount of smaller-scale structure missed by the polynomial fits that can only be observed using the individual antenna gradients, especially for the shortest baselines near the center of the array.

\section{Discussion}
Our exploration of a long, VHF observation of Cyg A with the VLA 
has successfully demonstrated the power of this instrument to 
characterize a variety of transient ionospheric phenomena.  For this 
observation, the typical $1 \sigma$ uncertainty in the $\delta \mbox{TEC}$ 
measurements was $3 \times 10^{-4}$ TECU, yielding more than an order of 
magnitude better sensitivity to TEC fluctuations than can be 
achieved with GPS-based relative TEC measurements 
\citep[see, e.g.,][]{her06}.\par
Large amplitude, long period waves reminiscent of MSTIDs are visible within the $\delta \mbox{TEC}$ data near dusk and dawn as well as other times intermittently throughout the night.  The polynomial-based approach we have detailed in \S 3.2 appears to be able to recover the properties of the full two-dimensional TEC gradients associated with these relatively large disturbances as they passed over the array.  This information can be used to estimate the size, speed and directions of such patterns down to scales of roughly half the size of the array ($\sim\! 20$ km).  This is demonstrated in more detail in a subsequent paper describing the spectral analysis of these data.\par
In addition, our approach of measuring the projected gradient at each antenna along each arm has shown that there are smaller-scale TEC fluctuations observed throughout the night, most prominently after midnight local time.  Thus, the VLA can be used to simultaneously study fine-scale ionospheric dynamics.  This may include a host of phenomena such as the small-scale distortions/structure within MSTID wavefronts, turbulent fluctuations from ion-neutral coupling within the lower ionosphere/thermosphere, and sporadic-E ($E_s$) layers.  In the case of $E_s$, \citet{cok09} demonstrated with a combination of VLA data and optical observations that many of the small-scale fluctuations seen by the VLA during summer nighttime are likely associated with $E_s$ layers.  \citet{cok09} showed that the TEC gradients caused by these layers are typically  $\approx \! 0.001$ TECU km$^{-1}$ which is easily detectable using the arm-based gradient method.  A specialized spectral analysis technique has also been developed for these measurements and will likewise be detailed in the companion manuscript to this paper.

\appendix
\section{Geometric Corrections}
\begin{figure}
\noindent\includegraphics[width=6in]{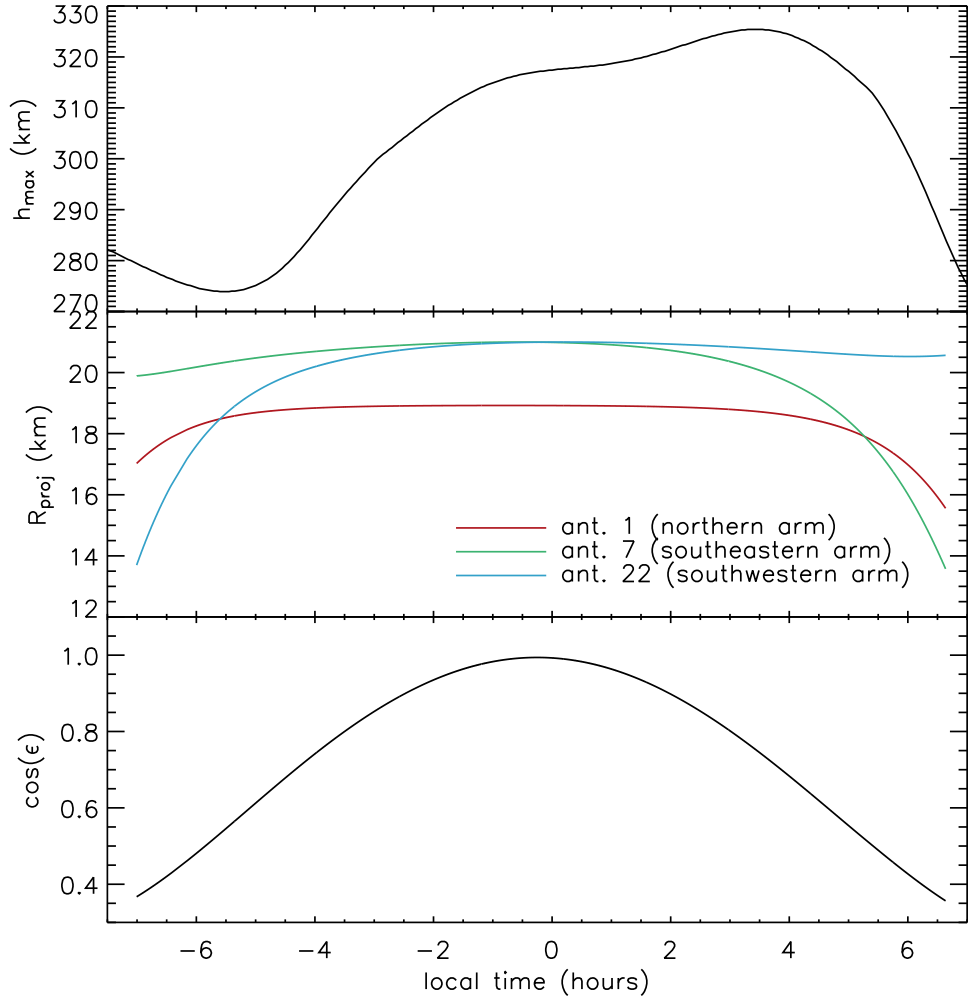}
\caption{Upper:  The height of the maximum electron density at the latitudes and 
longitudes of the ionosphere pierce-points during the 
observations as computed by the International Reference Ionosphere (IRI) software as a 
function of local time.  Middle:  For each of the three furthest antennas in the array, 
the distance of the ionosphere pierce-point of the antenna from that of the 
array center assuming a thin-shell model at the heights plotted in the upper panel 
as a function of time.  Lower:  The multiplicative factor used to correct the 
observed slant-$\delta \mbox{TEC}$ values assuming a thin-shell model at the heights 
plotted in the upper panel.}
\label{zion}
\end{figure}

Two basic geometric corrections must be applied to the antenna positions 
and $\delta \mbox{TEC}$ measurements so that they more accurately represent the 
actual conditions within the ionosphere.  Since we have used observations of 
Cyg A that include times when it is relatively close to the horizon, we cannot 
use a plane-parallel approximation.  Instead, we have used a thin shell 
approximation for the ionosphere where the shell is located at the height of 
maximum electron density, $z_{\mbox{\scriptsize ion}}$, as computed by the IRI software for the dates and times 
of the observations (see \S 3.1 and Fig.\ \ref{zion}).  
The full spherical corrections used are detailed below.\par
\begin{figure}
\noindent\includegraphics[width=6in]{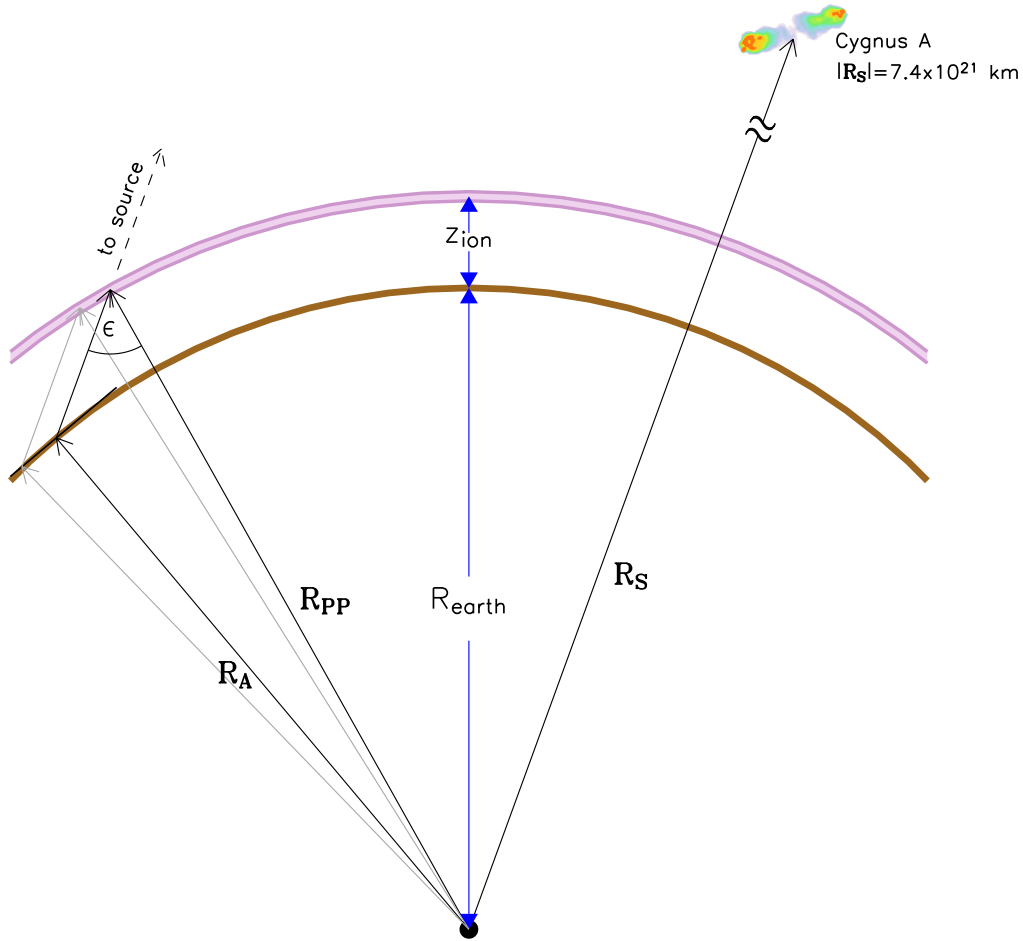}
\caption{A schematic representation (not to scale) of the procedure detailed here for computing the 
required geometric corrections.  See the text of this appendix for a detailed 
discussion of the different components of this schematic.  Note, the image used 
for Cyg A is a false-color image made with the VLA at 327 MHz \citep[see][]{laz06}.}
\label{geo}
\end{figure}

First, the positions of the antennas on the ground must be converted to projected 
positions within the ionosphere which, for a non-plane-parallel atmosphere, 
change with the elevation of the observed source.  For a spherical shell, we 
may define a ``pierce-point'' for each antenna where its line of sight to the 
source passes through the ionosphere.  The positions of these pierce-points 
relative to that for the center of the array can then be used as their projected 
ionosphere positions.  Fig.\ \ref{geo} provides a schematic illustration 
(not to scale) of how these positions are determined.  We first define a set three 
position vectors, $\mbox{\bf R}_{\mbox{\scriptsize \bf A}}$, $\mbox{\bf R}_{\mbox{\scriptsize \bf PP}}$, and 
$\mbox{\bf R}_{\mbox{\scriptsize \bf S}}$, which define the positions of the array center/antenna, 
the ionosphere pierce-point, and the observed source, respectively, with the 
center of the earth as the origin of the coordinate system.\par
Next, we note that 
the vast majority of astronomical sources, including Cyg A, are essentially 
infinitely far away, which implies that the line of sight from the array center/antenna 
location to the source is essentially parallel to that from the center of the 
earth to the sources, or 
$\mbox{\bf R}_{\mbox{\scriptsize \bf PP}}-\mbox{\bf R}_{\mbox{\scriptsize \bf A}} \mbox{ }||\mbox{ } \mbox{\bf R}_{\mbox{\scriptsize \bf S}}$.  If we define a ``left-handed'' coordinate system such that 
$\mbox{\bf R}_{\mbox{\scriptsize \bf A}}$ points along the $z$-axis, then the source 
position is given by
\begin{equation}
\frac{\mbox{\bf R}_{\mbox{\scriptsize \bf S}}}{|\mbox{\bf R}_{\mbox{\scriptsize \bf S}}|} = \mbox{cos}(h)\mbox{cos}(a) \hat{i} \nonumber  + \mbox{cos}(h)\mbox{sin}(a) \hat{j} + \mbox{sin}(h) \hat{k}
\end{equation}
where $h$ is the angular elevation of the source, $a$ is the azimuthal angle measured 
from north though east, and the $\hat{i}$ and $\hat{j}$ unit vectors point 
toward the north and east, respectively, as viewed from the array/antenna.  
Combining this with the assumption of parallel lines of sight to the source and 
the fact that the length of $\mbox{\bf R}_{\mbox{\scriptsize \bf PP}}$ is set to 
$\mbox{R}_{\mbox{\scriptsize earth}} + z_{\mbox{\scriptsize ion}}$ yields the 
following expression
\begin{equation}
\left ( \mbox{R}_{\mbox{\scriptsize earth}} + z_{\mbox{\scriptsize ion}} \right )^{2} = 
\left [ r \; \mbox{cos}(h)\mbox{cos}(a) + x_{A} \right ]^{2} \nonumber 
+ \left [ r \; \mbox{cos}(h)\mbox{sin}(a) + y_{A} \right ]^{2} \nonumber 
+ \left [ r \; \mbox{sin}(h) + z_{A} + \mbox{R}_{\mbox \scriptsize earth} \right ]^{2}
\end{equation}
where $x_A$, $y_A$ and $z_A$ are the coordinates of the antenna relative to the 
array center and $r=|\mbox{\bf R}_{\mbox{\scriptsize \bf PP}}-\mbox{\bf R}_{\mbox{\scriptsize \bf A}}|$.  
Since the antenna positions are known, $r$ is the only unknown variable.  Equation 
(A2) can then be rewritten as a quadratic equation and solved for $r$ keeping in 
mind that $0 \leq r < \mbox{R}_{\mbox{\scriptsize earth}} + z_{\mbox{\scriptsize ion}}$ 
which allows one to compute the $x$, $y$, and $z$ coordinates of 
$\mbox{\bf R}_{\mbox{\scriptsize \bf PP}}$ for the array center and each antenna in 
the current coordinate system.  Following this, a coordinate rotation was performed 
such that $\mbox{\bf R}_{\mbox{\scriptsize \bf PP}}$ for the array center pointed 
along the $z$-axis and the $x$ and $y$ axes pointed toward north and east, respectively, 
as viewed from the location on the earth directly below the array center pierce-point.  These rotated 
coordinates were then taken to be the $x$ and $y$ antenna positions projected onto the 
ionosphere thin shell for each time step.  Fig.\ \ref{geo} shows a graphical 
representation of these computations for the array center (in black) and for an 
exemplar antenna (in gray).\par 
The second correction deals with the fact that the path length through the ionosphere 
is increased when the observed 
source is closer to the horizon.  For a thin spherical shell, it is increased by a 
factor of $\mbox{sec}(\epsilon)$ where $\epsilon$ is the angle between the line of 
sight from the VLA to the source and a line from the ionosphere pierce-point to the 
location on the earth directly below it.  In the schematic in Fig.\ \ref{geo}, $\epsilon$ is the angle 
between the position vectors $\mbox{\bf R}_{\mbox{\scriptsize \bf PP}}$ and 
$\mbox{\bf R}_{\mbox{\scriptsize \bf PP}}-\mbox{\bf R}_{\mbox{\scriptsize \bf A}}$.  
Therefore, to compute the factor needed to correct our $\delta \mbox{TEC}$ measurements, 
$\mbox{cos}(\epsilon)$, we simply computed the dot product between these two vectors 
and divided by the product of their lengths, 
$r(\mbox{R}_{\mbox{\scriptsize earth}} + z_{\mbox{\scriptsize ion}})$.\par
\begin{figure}
\noindent\includegraphics[width=6in]{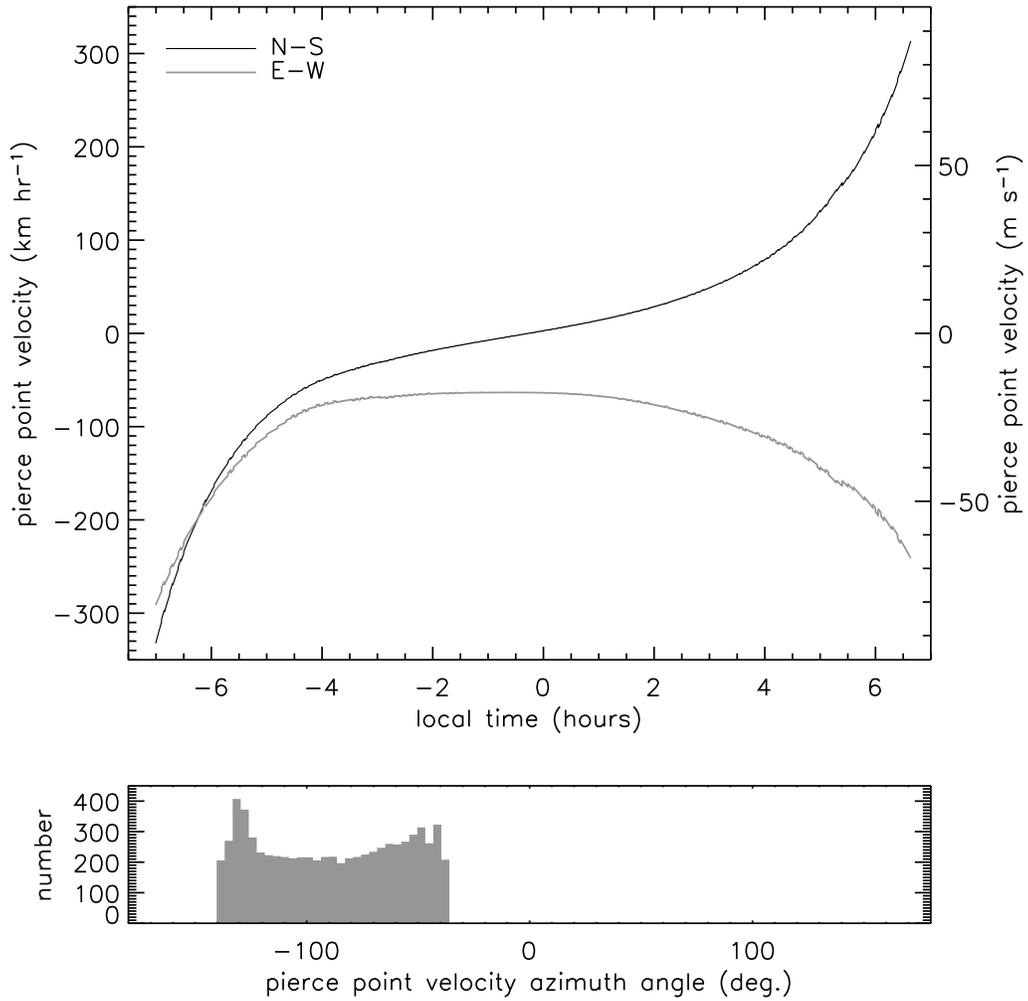}
\caption{Upper:  The estimated apparent velocity of Cyg A at the location within the 
thin-shell ionosphere of the array center pierce-point in both the north-south (black) 
and east-west (gray) directions as functions of time.  Lower:  A histogram for the 
distribution of azimuth angles measured from north through east for the velocity 
vectors plotted in the upper panel.}
\label{svel}
\end{figure}

Finally, while computing the above geometric corrections, we also computed estimates 
of the apparent motion of Cyg A within the coordinate system of each time step.  This 
was done to estimate the degree of Doppler shifting of the temporal/spatial 
frequencies of any detected wave phenomena.  We did this 
for each time step by recomputing the position of the array center pierce-point 
for the two nearest time steps within the coordinate system of the current time step.  
These positions were then used to numerically compute the time derivatives of the 
$x$ and $y$ coordinates of the array center pierce-point to obtain the north-south 
and east-west components of the sidereal velocity.  These are plotted in Fig.\ 
\ref{svel} as functions of time along with a histogram for the azimuth angle 
(measured north through east) of the sidereal velocity vector for the entire observing 
run.  One can see from this figure that the velocities were sometimes significant, 
especially 
when the source was at lower elevations.  In addition, while the motion is generally from 
east to west, as one would naively assume, there is a significant spread in position 
angle of more than $100^{\circ}$.

\begin{acknowledgments}
The authors would like to thank the referees for useful comments and suggestions.  Basic research in astronomy at the Naval Research Laboratory is supported 
by 6.1 base funding.  The VLA was operated by the National Radio Astronomy Observatory which is a facility 
of the National Science Foundation operated under cooperative agreement by 
Associated Universities, Inc.  Part of this research was carried out at the Jet Propulsion Laboratory, California Institute of Technology, under a contract with the National Aeronautics and Space Administration.
\end{acknowledgments}

\end{article}

\clearpage

\clearpage

\clearpage

\clearpage

\clearpage

\clearpage

\clearpage

\clearpage

\clearpage

\clearpage

\clearpage

\clearpage

\end{document}